\documentclass[12pt]{article}

\usepackage{ifthen}
\usepackage{times}

\usepackage{cellspace, tabularx, booktabs}
\addparagraphcolumntypes{X}

\newenvironment{sciabstract}{%
\begin{quote} \bf}
{\end{quote}}

\usepackage{tcolorbox} 
\usepackage{graphicx}
\usepackage[english]{babel}
\usepackage[utf8]{inputenc}
\usepackage{lineno}
\usepackage{amsmath, amssymb}
\usepackage{hyperref}
\renewcommand{\emph}[1]{{\it #1}}

\def\fig#1{Fig.~\ref{fig:#1}}
\def\sfig#1#2{\fig{#1}#2}

\def\tatm{T_a}
\def\tsrf{T_s}
\def\sens{H}
\def\flat{LE}

\def\trad{T_{\mbox{\footnotesize rad}}}

\def\lw{LW}
\def\flw{F_{\lw}^\downarrow}

\def\sw{SW}
\def\fsw{F_{\sw}^\downarrow}

\def\sfiglab#1{{\bf #1}}
\def\trad{T_{rad}}

\def\pw{p_a}
\def\ps{p_s}

\def\Hin{H_{in}}

\def\tw{T_w}

\def\uw{U_w}

\definecolor{py1}{HTML}{1f77b4}
\definecolor{py2}{HTML}{ff7f0e}
\definecolor{py3}{HTML}{dbb40c} 
\definecolor{py4}{HTML}{2ca02c} 
\definecolor{py5}{HTML}{9467bd}
\definecolor{py6}{HTML}{8c564b}
\definecolor{py7}{HTML}{e377c2}
\definecolor{py8}{HTML}{7f7f7f}
\definecolor{py9}{HTML}{d62728}

\colorlet{rwalk}{gray!80!black}
\colorlet{warm}{py9}
\colorlet{midl}{py4}
\colorlet{cold}{py1}

\graphicspath{{figures/}}

\title{The ``teapot in a city'': a paradigm shift in urban climate modeling}

\author{
{Najda} {Villefranque}$^\ast$,$^{1,2,3}$
{Frédéric}   {Hourdin},$^{1}$
{Louis}      {d'Alençon},$^{1}$\\
{Stéphane}   {Blanco},$^{2}$
{Olivier}    {Boucher},$^{1}$
{Cyril}      {Caliot},$^{4}$\\
{Christophe} {Coustet},$^{5}$
{Jérémi}     {Dauchet},$^{6}$
{Mouna}      {El Hafi},$^{7}$\\
{Vincent}    {Eymet},$^{5}$
{Olivier}    {Farges},$^{8}$
{Vincent}    {Forest},$^{5}$\\
{Richard}    {Fournier},$^{2}$
{Jacques}    {Gautrais},$^{9}$
{Valéry}     {Masson},$^{3}$\\
{Benjamin}   {Piaud},$^{5}$
{Robert}     {Schoetter},$^{3}$\\
\\
\normalsize{$^1$LMD/IPSL/SU, CNRS, Paris, {75005}, {France},}\\
\normalsize{$^2$Laplace, INP/Université de Toulouse/CNRS, Toulouse, France,}\\
\normalsize{$^3${CNRM}, {Université de Toulouse, Météo-France, CNRS}, Toulouse, France}\\
\normalsize{$^4$LMAP, CNRS, UPPA, E25, Anglet, France,}\\
\normalsize{$^5$Méso-Star, Longages, France,}\\
\normalsize{$^6${Institut Pascal}, Université Clermont Auvergne, Clermont Auvergne INP, CNRS, Clermont-Ferrand, France,}\\
\normalsize{$^7${Centre RAPSODEE}, Université de Toulouse, Mines Albi, UMR CNRS 5302, {Campus Jarlard}, {Albi}, {France},}\\
\normalsize{$^8${LEMTA}, Université de Lorraine, CNRS, {Nancy}, {France},}\\
\normalsize{$^9${CRCA, CBI}, {Université de Toulouse, CNRS}, Toulouse, France,}\\
\\
\normalsize{$^\ast$To whom correspondence should be addressed; E-mail: najda.villefranque@lmd.ipsl.fr}
}

\begin{document}
\baselineskip24pt
\maketitle

Revisiting Feynman-Kac's path integrals using computer graphics ray-tracing to
anticipate the consequences of global warming.

\begin{sciabstract} 
  Urban areas are a high-stake target of climate change mitigation and
  adaptation measures. To understand, predict and improve the energy
  performance of cities, the scientific community develops numerical models
  that describe how they interact with the atmosphere through heat and moisture
  exchanges at all scales. In this review, we present recent advances that are
  at the origin of last decade's revolution in computer graphics, and recent
  breakthroughs in statistical physics that extend well established
  path-integral formulations to non-linear coupled models. We argue that this
  rare conjunction of scientific advances in mathematics, physics, computer and
  engineering sciences opens promising avenues for urban climate modeling and
  illustrate this with coupled heat transfer simulations in complex urban
  geometries under complex atmospheric conditions. We highlight the potential
  of these approaches beyond urban climate modeling, for the necessary
  appropriation of the issues at the heart of the energy transition by
  societies.
\end{sciabstract}

\section*{Introduction}

In the face of global warming, scientists are urged to provide climate
information to support mitigation and adaptation policies. To address this
challenge, new fields of research have emerged that aim at filling the gap
between climate change projections and societal needs. Progress is slow due to
the complexity of the systems that need to be analyzed to provide relevant
climate information to end users. The questions are multidisciplinary hence the
expertise of a wide range of communities from climate to human sciences must be
involved. The models that are needed to predict the effects of climate change
must account for a wide variety of processes characterized by large ranges of
spatial and temporal scales. They must be able to ingest large amounts of data
from local constraints to climatic records of time-varying meteorological
conditions. Uncertainties related to each component must be quantified and
propagated through the various model layers. Cities are a high-stake target of
adaptation policies, and an archetype of such complex systems.

The prime effect of urbanization on the local climate, investigated since the
1980s, is known as the Urban Heat Island (UHI) effect: cities are almost always
warmer than their environment \cite{oke_city_1973, arnfield_two_2003}. The
resulting heat stress, intensified by global warming, leads to health
impairment, increased mortality \cite{Gosling_2009, Gabriel_2011} and/or an
increase in energy consumption for air conditioning which positively feedbacks
on the UHI \cite{grimmond_urbanization_2007, Munck_2013}. As more than half of
the world's population now lives in urban areas \cite{UN_2019}, it has become
crucial to adapt cities and design new ones in a way that both improves thermal
comfort and reduces energy consumption \cite{grimmond_integrated_2020,
baklanov_urban_2018}.  Climate change mitigation and adaptation measures range
from home improvement and renovation by owners, climate-proof building design
by architects \cite{Akbari_2016}, use of new materials and {urban cooling
technologies} \cite{santamouris_passive_2016}, introduction of urban vegetation
\cite{gunawardena_utilising_2017}, and exploitation of the surrounding
landscape potential \cite{Masson_2020} by urban planners. Identifying and
developing new \emph{urban cool islands} has become a priority in some cities.
To fulfill this objective, international organizations such as the World
Meteorological Organization advocate for the development of \emph{climate
services}, through which climate scientists are expected to deliver
``high-quality, science-based climate information tailored to city requirements
to improve urban resiliency and to support the sustainable development of the
cities in the world'' \cite{baklanov_urban_2018}.

A classical approach to model climate-related impacts for urban climate
services is to rely on either statistical models or urbanized atmospheric
models. Accounting for the detailed city geometry, the heterogeneity of urban
materials, and the variety of physical processes occuring over a wide range of
scales is however extremely challenging \cite{Oke_2017}. Numerous complementary
approaches exist and range from large-scale physical models that account for
climate change but drastically simplify the urban geometry \cite{Jacob_2014,
Martilli_2002, Masson_2000, Kikegawa_2003, Salamanca_2009, Bueno_2012,
Garuma_2018}, to building-resolving models that account for the detailed
features of the city but are limited to either small-domain simulations
conducted over short time periods \cite{Bruse_1998, Maronga_2020, Crawley_2001,
Vorger_2014, Robinson_2009, Vinet_2000, Miguet_2002}, or to current climate
conditions for statistical models trained on observational datasets
\cite{lindberg_urban_2018}. A brief review of these approaches can be found in
Table~\ref{table:box}.

\begin{table}
\caption{Strengths and limitations of existing urban climate models}
\label{table:box}
    \begin{tabularx}{\linewidth}{>{\hsize=.95\hsize}SX}
    \toprule \centerline{Physical atmospheric models with parameterized urban canopies} \\ \midrule 
    Transient (e.g., from 1970 to 2100) global or regional climate simulations
      can be made at 10 to 100~km resolution using atmospheric models
      \cite{Jacob_2014}. Atmospheric models at hectometric to kilometric
      resolution can provide simulations for a few days and up to one year
      \cite{Skamarock_2008, Lac_2018}). In both cases, the cities cannot be
      represented explicitly. Rather, urban canopy models like the Building
      Effect Parametrisation BEP \cite{Martilli_2002} or the Town Energy
      Balance TEB \cite{Masson_2000} (including a building energy model
      \cite{Kikegawa_2003, Salamanca_2009, Bueno_2012}) are used to estimate
      the effect of subgrid radiative and heat transfer on the air temperature,
      winds and water balance.  The geometry of the city is greatly simplified,
      usually using the ``urban canyon'' approximation (an infinite street with
      two facing walls \cite{Garuma_2018}). These models provide useful
      information such as the impact of the urban heat island on the building
      energy consumption, sometimes in an operational service such as in
      Beijing \cite{He_2019}.  However, they cannot provide information at the
      scale of a flat or a building nor do they help to assess the impact of
      small-scale adaptation measures. ~\\
    \midrule \centerline{Physical building-resolving models with parameterized environment} \\ \midrule 
    Higher resolution building-resolving micrometeorological models can
      represent the detailed urban geometry but simulations are limited to a
      neighbourhood (typically 500~m $\times$ 500~m) and simulations up to a
      few days \cite{Bruse_1998, Maronga_2020}. Building energy models like
      EnergyPlus \cite{Crawley_2001} simulate the energy budget of an
      individual building accounting for a high level of detail (e.g., room
      allocation, building occupant behaviour \cite{Vorger_2014}, types of
      shading elements...). They rely on other models such as CITYSIM
      \cite{Robinson_2009} or SOLENE \cite{Vinet_2000, Miguet_2002} to model
      the environmental effects, like the shading of adjacent buildings. In
      this approach, investigation of the impact of climate change is severly
      limited by the difficulty to handle meteorological forcings.\\
    \midrule \centerline{Statistical models} \\ \midrule 
      They are usually trained on local observations and limited to the site
      and conditions under which observations are available. They sometimes use
      statistical laws calibrated on various sites to provide estimations of
      quantities on other neighborhoods (e.g., the Urban Multi-scale
      Environmental Predictor UMEP \cite{lindberg_urban_2018}). They are
      computationally efficient but assume constant statistical relationships
      between historical and future climate. They are limited to the resolution
      of the observational data, although the impacts of processes occurring at
      all scales are inherently integrated into the measurements that
      constitute the training dataset.~\\
      \bottomrule
  \end{tabularx}
\end{table}

\begin{figure*}\centering
  \includegraphics[width=\textwidth]{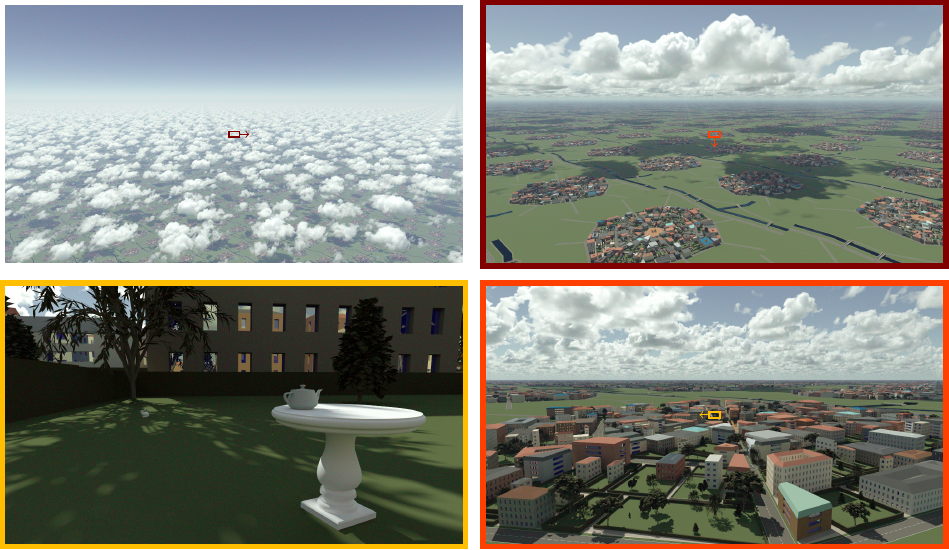}
  \caption{The teapot in a city under cumulus clouds, in reference to the
  ``teapot in the stadium'' problem. The four pictures are sampled from an
  animated movie
  (\href{https://www.lmd.jussieu.fr/~nvillefranque/pages/teapot_city}{\url{https://www.lmd.jussieu.fr/~nvillefranque/pages/teapot_city}})
  we produced using the htrdr model \cite{villefranque_path-tracing_2019}
  (\href{https://www.meso-star.com/projects/htrdr/htrdr.html}{\url{https://www.meso-star.com/projects/htrdr/htrdr.html}})
  that solves radiative transfer in the atmosphere and in cities. Each image
  features a different cloud field, camera
  and sun positions. Periodic conditions were used for the city geometry and
  the cloud fields to demonstrate insensitivity to the scene dimension. Cities
  and cloud fields of larger extent can be rendered with open boundary
  conditions as easily, provided that the data is available. The urban geometry
  was generated using a procedural generator
  (\href{https://gitlab.com/meso-star/city\_generator}{https://gitlab.com/meso-star/city\_generator})
  based on sampling distributions that represent the buildings characteristics
  (height, spacing...) and various tree geometries. The spectrally varying
  radiative properties of the materials were taken from the Spectral Library of
  Impervious Urban Materials (SLUM) database \cite{kotthaus_derivation_2014}.
  The cloudy atmosphere was simulated using the Meso-NH Large-Eddy Simulation
  (LES) model \cite{lafore_meso-NH_1997, Lac_2018} and represents a typical
  fair-weather cumulus field evolving over a flat ground
  \cite{brown_large_2002} at 8 meter resolution on a 15 $\times$ 15 $\times$ 4
  km$^3$ domain with horizontally periodic boundary conditions with 3D fields
  output every 15 seconds between 11:30 and 13:00 Local Solar Time (LST).
  }
  \label{fig:nuages}
\end{figure*}

In this study, we present a new paradigm for  multi-scale modeling of coupled
radiative and heat transfer in complex urban geometry under changing climate.
It relies on probabilistic models solved by Monte Carlo methods and builds on
recent advances in computer graphics. The ``teapot in a stadium'' problem
\cite{haines_spline_1988}, namely the difficulty to render small-scale details
(the teapot) within a very large scene (a stadium), has been solved
\cite{wald_embree_2014}. The computing time associated with path-tracing in 3D
scenes {is now close to} insensitive to the scene complexity. As Monte Carlo
methods used to solve the radiative transfer equation are independent from the
description of the geometric data, increasing the computation accuracy can be
achieved through improving the physical model or improving the data in
completely independent developments. In the world of 3D animation for film
production, this property has freed up the artists who do no longer have to
compromise on the complexity of their scenes to comply with the limitations of
rendering algorithms. Similarly computer scientists have been able to include
more complex physics in their algorithms, producing ever more realistic images
by taking into account every detail of the virtual scene in a physically
consistent manner. We illustrate this in \fig{nuages} with four images sampled
from the animated movie of a ``teapot in a city under cumulus clouds'',
available at
\href{https://www.lmd.jussieu.fr/~nvillefranque/pages/teapot_city}{\url{https://www.lmd.jussieu.fr/~nvillefranque/pages/teapot_city}}.
It is based entirely on physical principles, both for simulating clouds and for
light propagation.

We envision that the exact same framework of formulating physical processes as
path integrals and integrating them numerically with Monte Carlo path-tracing
methods could lead to a similar revolution in urban modeling. Here we review
recently published results which, put together, allow for  computations that
were previously unthinkable. Before reviewing these breakthroughs and
reflecting on the perspectives that are opening up for urban climate services,
we present the foundations of these methods using a very simple example of
computing the energetics of a two-dimensional building.  In doing so, we hope
to introduce the readers to the key concepts of the framework, from the most
technical aspects to their most profound implications, and illustrate the
benefits for analysing and understanding complex systems.

\section*{A simple example of path-integral formulation}

\begin{figure*}\centering
  \includegraphics[width=\textwidth]{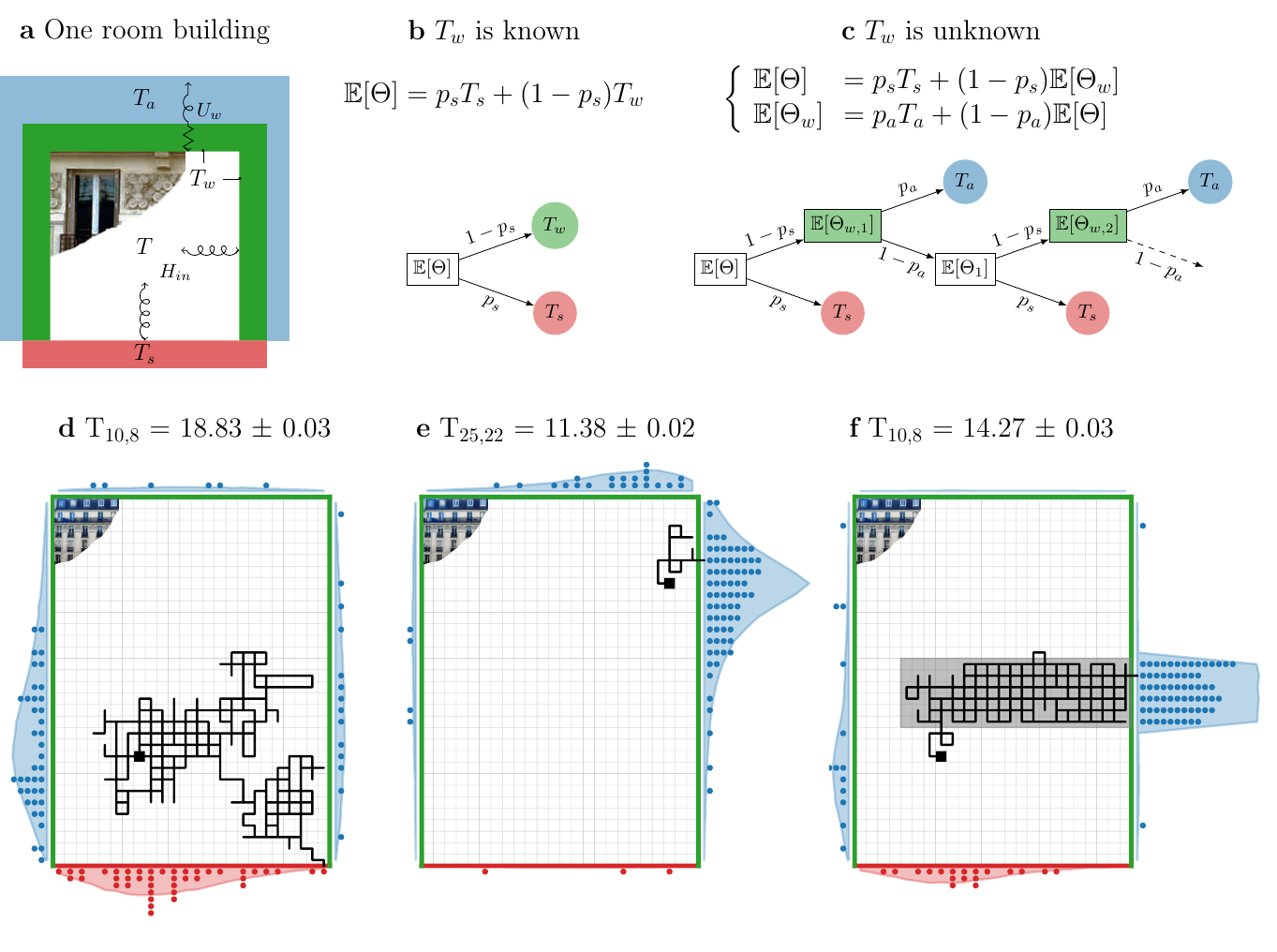}
  \caption{Heat transfers in 2D buildings of \sfiglab{a-c} one room and
   \sfiglab{d-f} $N \times M$ rooms. \sfiglab{a}, $T$ is the temperature of the
   room's perfectly-mixed air, $\tsrf$ the temperature of the ground floor,
   $\tw$ the temperature of the three other walls, $\tatm$ is the temperature
   of the environmental perfectly-mixed air. Heat exchange between the inside
   air and the interior walls is driven by convection, with convective thermal
   conductance (CTC) $\Hin$. Heat exchange between the interior walls and the
   outside air is driven by conduction in the wall and convection outside, of
   global thermal conductance $\uw$. \sfiglab{b}, $T$ is the average
   of $\tw$ and $\tsrf$, which is also the expectation of $\Theta$ whose
   outcomes are $\tsrf$ with probability $\ps$ and $\tw$ with probability
   $1-\ps$. \sfiglab{c}, $\tw$ is itself the expectation of $\Theta_w$. One
   realization of $\Theta$ is sampled by first sampling $\Theta_{w,n}$ and then
   $\Theta_{n}$ successively until an outcome ($\tsrf$ or $\tatm$) is found.
   $(\Theta_{w,n})_{n=1,..N}$ and $(\Theta_{n})_{n=1,..N}$ are collections of
   independent and identically distributed random variables that have the same
   probability law as $\Theta_w$ and $\Theta$ respectively.  \sfiglab{d-f},
   $T_{i,j}$ is the temperature in room $(i,j)$ (black square).  The exterior
   walls have the same properties as in \sfiglab{a}, $\tatm=10^\circ$C,
   $\tsrf=30^\circ$C. The interior walls all have the same CTCs except in the
   gray zone of \sfiglab{f} where they are a hundred times larger, which is
   symptomatic of a thermal bridge.  The first sampled path (black line), the
   end locations of the first 100 sampled paths (blue and red points) and the
   distribution of the end locations of the 100,000 sampled paths (blue and red
   shadings) are shown for each simulation.
   }
   \label{fig:couplage} 
\end{figure*}

The framework we present here is built on two fundamental ideas: the
formulation of deterministic physical models as integrals over path spaces
\cite{feynman2005space, kac1951some}, and double randomization
\cite{sabelfeld_monte_1991}. Let us illustrate them for a simple 2D model of
the steady-state temperature $T$ of perfectly-mixed air inside a square room
framed by three segment walls (temperature $\tw$) and ground floor (temperature
$\tsrf$), surrounded by air at temperature $\tatm$ (see schematic in
\sfig{couplage}{a}). Per analogy with an electrical network, $T$ can be written
as the average of the ground and walls' temperature weighted by the wall
convective thermal conductances (convective coefficients times length of the
wall). If the room is a square and the convective coefficient is constant then
$T = \ps \tsrf + (1-\ps)\tw$ with $\ps=1/4$.

Let us shed a probabilistic light on this deterministic problem and interpret
$T$ as the expectation of a random variable $\Theta$ following a Bernoulli's law
of parameter $\ps$, with outcomes $\tsrf$ and $\tw$ (see \sfig{couplage}{b}).
An unbiased estimate of $T$ can then be produced by averaging
a large number of realizations of $\Theta$. More generally, whenever it is
possible to formulate a quantity as an expectation of a (function of) discrete
or continuous random variable(s), then this quantity can be estimated using
Monte Carlo methods. This is the first fundamental idea of the framework.

Now let the value of $\tw$ be unknown. Defining a global thermal wall
conductance to account for conduction in the wall and convection oustide, and
considering that heat fluxes are continuous at the inner wall surface, $\tw$ can again
be written as the expectation of a random variable following another
Bernoulli's law of parameter $\pw$, with possible outcomes $\tatm$ and $T$ (see
\fig{couplage}c).

Combining the expressions for $T = \mathbb{E}[\Theta]$ and $\tw =
\mathbb{E}[\Theta_w]$  yields a recursive expression. Most of the time, there
is no closed-form equivalent for the recursive expressions that come from the
probabilistic formulation of a deterministic problem, therefore the ``global''
law that directly gives the probability of the final outcomes ($\tatm$ or
$\tsrf$) cannot be sampled. This is where the concept of double randomization
is needed. It consists in sampling the ``local'' probability laws (of $\Theta$
and $\Theta_w$) successively until finding an outcome ($\tatm$ or $\tsrf$) for
the random sequence or \emph{path}. This is illustrated in \sfig{couplage}{c}.
The justification for double randomization is mathematically trivial (it comes
from the law of total expectation), albeit conceptually subtle.  This second
fundamental idea explains why Monte Carlo methods are insensitive to the
problem's dimension: each step of the sampling procedure is entirely oblivious
to the rest of the model.

Now let the building consist of many rooms $(i,j)$ at temperature
$T_{i,j}=\mathbb{E}[\Theta_{i,j}]$ (see \sfig{couplage}{d-f}) with
known thermal conductances and boundary conditions ($\tsrf$ and $\tatm$).
Implementing the same strategy as before to compute $T_{i,j}$ yields a Monte
Carlo algorithm that consists of successively sampling neighboring rooms,
starting at $(i,j)$, until finding an outcome (a boundary condition). $T_{i,j}$
is then estimated as the mean outcome. One realization of $\Theta_{i,j}$ can be
represented as a path throughout the building from room $(i,j)$ to the location
of the outcome, as in \sfig{couplage}{d-f}. Randomly constructing these paths
step by step using double randomization ensures that the two possible outcomes
are sampled in the correct proportions. \fig{couplage}{d-f} displays the
distribution of the path outcomes; more paths end in the neighborhood of the
$(i,j)$ room than at opposite walls, except in \fig{couplage}{f} where the
paths show that most of the heat is lost through the poorly insulated part of
the building. 

The building heat loss is proportional to the difference between the outside
air and the average temperature of the boundary rooms of the building.
Estimating this average temperature instead of the temperature of one
particular room can be done using the same algorithm, except that instead of
fixing $(i,j)$ beforehand, a starting room is randomly sampled at the beginning
of each path. As ``probe'' type computations do not rely on solving the entire
field, the quantity is obtained at a point or on average for approximately the
same computing time.

\section*{Breakthroughs in Monte Carlo methods}%

A major feature of Monte Carlo methods is their insensitivity to the dimension
of the problem. In the 2D building of \fig{couplage}, computing the heat loss
of the $N \times M$ room building at 1\% precision takes roughly the same
computing time as computing the heat loss of a building twice as high, made of
twice as many storeys. This feature has been central in the use of these
methods throughout many scientific fields ever since Nicholas Metropolis and
Stanislaw Ulam coined the name ``Monte Carlo Method'' in their famous 1949
article \cite{metropolis1949monte}. It is the same feature that makes Monte
Carlo methods so powerful to solve recursive equations such as the Fredholm
equations of the second kind, and extensively used in particle transport from
neutronics to rarefied gas to radiative transfer \cite{lux2018monte}. {After
use of kernel iterative method, the Fredholm equation admits Von Neumann series
representation making the integration problem of infinite dimension
\cite{farnoosh2008monte, doucet2010solving},} but double randomization simply
translates it into successive sampling collision events one after the other
(much as successive neighbouring rooms are sampled in the example of the 2D
building) until an outcome is found.  A major step was achieved when Monte
Carlo methods were extended to problems that were not initially formulated into
the framework of statistical physics. This led to important advances in the
field of applied mathematics where Monte Carlo methods are now routinely used
for large matrices inversion, and in physics when Richard Feynman and Marc Kac
formulated the general solution of the differential equations that model
advecto-reacto-diffusive processes as the expectation of a Wiener process
\cite{del2004feynman,
ito2012diffusion,kac1951some,feynman2005space,kac1947random}. This opened new
fields of application with for instance Monte Carlo simulations of Brownian
motions to solve three-dimensional transient diffusion
\cite{lapeyre2003introduction, muller1956some, haji1967solution,
sabelfeld_global_2019}.

However, the insensitivity to the dimension was lost when problems included
either (i) 3D geometries characterized by wide ranges of scales, (ii) coupled
models of different natures or (iii) nonlinearities. The three following
breakthroughs could overcome these limitations:

\begin{figure*}\centering
  \includegraphics[width=\textwidth]{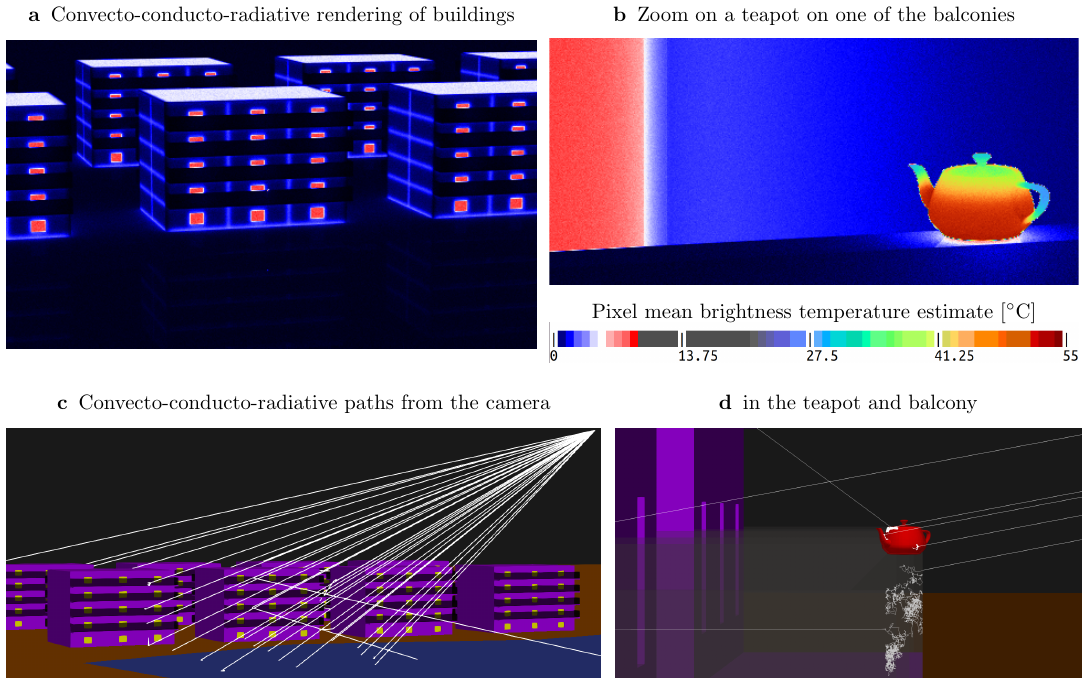}
  \caption{Physical infrared rendering of 3D buildings near a lake, in steady
  state, at night. The brightness temperature equivalent to the radiation
  emitted by the buildings, ground and atmosphere and received at the virtual
  camera is computed in each pixel by solving detailed heat transfers in the
  scene, using the Stardis software
  (\url{http://meso-star.com/projects/stardis/stardis.html}). Paths start at
  the camera; conduction is simulated using $\delta$-sphere walks inside the
  solids; radiative exchanges are sampled between surfaces. Paths stop upon
  reaching a boundary condition: the temperature of the atmosphere
  (0$^\circ$C), and rooms (20$^\circ$C) by convection or the brightness
  temperature of the atmosphere (0$^\circ$C) by radiation. They can also stop
  in the teapot which contains water at an imposed temperature of 60$^\circ$C.
  \sfiglab{a,b}, results of convective-conductive-radiative Monte Carlo
  simulations for two views: \sfiglab{a}, a few buildings and \sfiglab{b}, a
  zoom on the teapot. Note that in \sfiglab{a}, the teapot is already on the
  first floor balcony of the middle building; it increases the mean brightness
  temperature of one of the pixels inside the red frame. \sfiglab{c,d}, 3D
  visualization of the scene and of a few paths sampled during the simulations.
  The scene consists of 33,958 facets (10,234 facets to describe the teapot and
  23,724 for the buildings). Each image consists of 480$\times$280 independent
  Monte Carlo estimates (one per pixel, 512 paths each).
  }
  \label{fig:benjamin} 
\end{figure*}

Breakthrough (i) was achieved by the computer graphics community who has
invested and revolutionized the field of Monte Carlo physically-based rendering
\cite{Kajiya1986, Cook1984, veach_robust_1998, pharr_physically_2018}. In order
to increase the realism of animated movies they increased the geometric details
of the rendered virtual scenes, thus increasing the number of facets to be
tested for intersection when tracing paths. That led them to conceive
hierarchical structures to organize the data in memory so that the cost of
path-tracing became independent of the number of facets describing the scene
\cite{clark_hierarchical_1976, kay_ray_1986, glassner_introduction_1989,
wald_embree_2014}. Until recently however, ray tracing procedures were
inefficient in highly heterogeneous media such as clouds, due to the
non-linearity of Beer's exponential law. A first step toward efficiency was to
make ray tracing independent from the description of the medium with
null-collision algorithms \cite{raab_unbiased_2006, galtier_integral_2013}.
This opened new possibilities that begun to be investigated in physics
\cite{eymet_null-collision_2013, galtier_radiative_2016, galtier_symbolic_2017,
sans_null-collision_2021} and computer graphics \cite{novak2018monte,
miller19null, georgiev19integral}. From there, the hierarchical structures were
extended to handle complex volumetric data
\cite{villefranque_path-tracing_2019}, thereby making the cost of numerical
computations in cloudy atmosphere insensitive to the details of the cloud
description. This first breakthrough is illustrated in \fig{nuages}. In each
image, all the details of the clouds and the city, including the teapot, are
taken into account, even when they are not perceivable to the eye.

Breakthrough (ii) was to understand that the double randomization concept
enables the coupling of models with no theoretical limit to the number, nature
or scale of the represented processes \cite{fournier_radiative_2016}. As an
illustration, \fig{benjamin} displays a virtual infrared image of buildings at
night time, that was rendered by tracing paths from a camera solving a coupled
conductive-convective-radiative equation. {The continuity of the boundary
fluxes} is reformulated as a probability to switch from one transfer mode to
the other, much as in the example of the 2D building. In this example, {the
temperature of the rooms in the building} is 20$^\circ$ and the outside air
temperature is
0$^\circ$. Thermal losses through windows, roofs and floors are captured by the
infrared virtual camera (panel a) by explicitly simulating heat transfer
processes in the detailed geometry. A few brighter pixels at the center of the
image are due to the presence of a hot teapot set on the balcony, as revealed
by the zoomed image of panel b. The teapot is half filled with hot liquid,
creating a temperature gradient on the teapot surface, with warm bottom and
colder lid, pout and handle. These fine-scale details were captured by the
paths simulated inside the teapot geometry, with some of the paths also
exploring parts of the geometry that are not directly visible to the camera,
capturing the larger scale transfers shown in panel a.  The teapot is {an
iconic} object from the comptuer graphics community, it has no climatic
relevance but here serves to demonstrate that large ranges of scales can be
seamlessly integrated. The thermal loss integrated over the entire city can be
computed in the same framework; since the temperature field never needs to be
estimated, the computational expense will be approximately the computing time
associated with one pixel of the images. Other coupled models have been solved
based on the same idea, such as the radiative transfer equation coupled with a
spectroscopy model, directly integrating the spectral lines, thereby avoiding
the heavy precomputation of absorption coefficient spectra
\cite{galtier_radiative_2016}.  It was also used in process engineering to
solve a cascade of embedded models from radiative transfer to electromagnetism
to thermokinetic coupling to spectroscopy, in order to estimate the biomass
production of a photoreactor system at the industrial scale
\cite{gattepaille_modeles_2020}. The paths are no longer restricted to the 3D
space but ``travel'' through models of different natures.

Although ``Direct Simulation Monte Carlo'' methods \cite{bird_approach_1963}
have for long addressed non-linear physics by simultaneously tracing large
numbers of paths so that the tracked particles could interact with each other,
they are fundamentally sensitive to the model or domain dimensions: increasing
the accuracy of the estimates implies increasing the particle density
everywhere in the spatial domain and the other domains of integration of the
problem. Recent advances have paved the way for nonlinear Monte Carlo
calculations that preserve the insensitivity property (breakthrough (iii)).
First, and although atmospheric radiative transfer is fundamentally linear, the
null-collision approach mentioned above can be seen as a way to bypass Beer's
non-linearity \cite{elhafi_three_2020}.  It has been shown that other types of
non-linearities could be treated using ramified paths
\cite{dauchet_addressing_2018, terree_addressing_2022}. The algorithms are more
complex but the paths can still be sampled independently, thus preserving all
the benefits of probe Monte Carlo approaches. Further investigations have shown
that iterative methods could be used to limit the recursivity level of the path
ramifications, enhancing the practicability of the method
\cite{tregan_thermique_2020}.

\section*{Implications for urban climate services}%
\label{sec:climat}

\begin{figure*}
	\includegraphics[width=.95\textwidth]{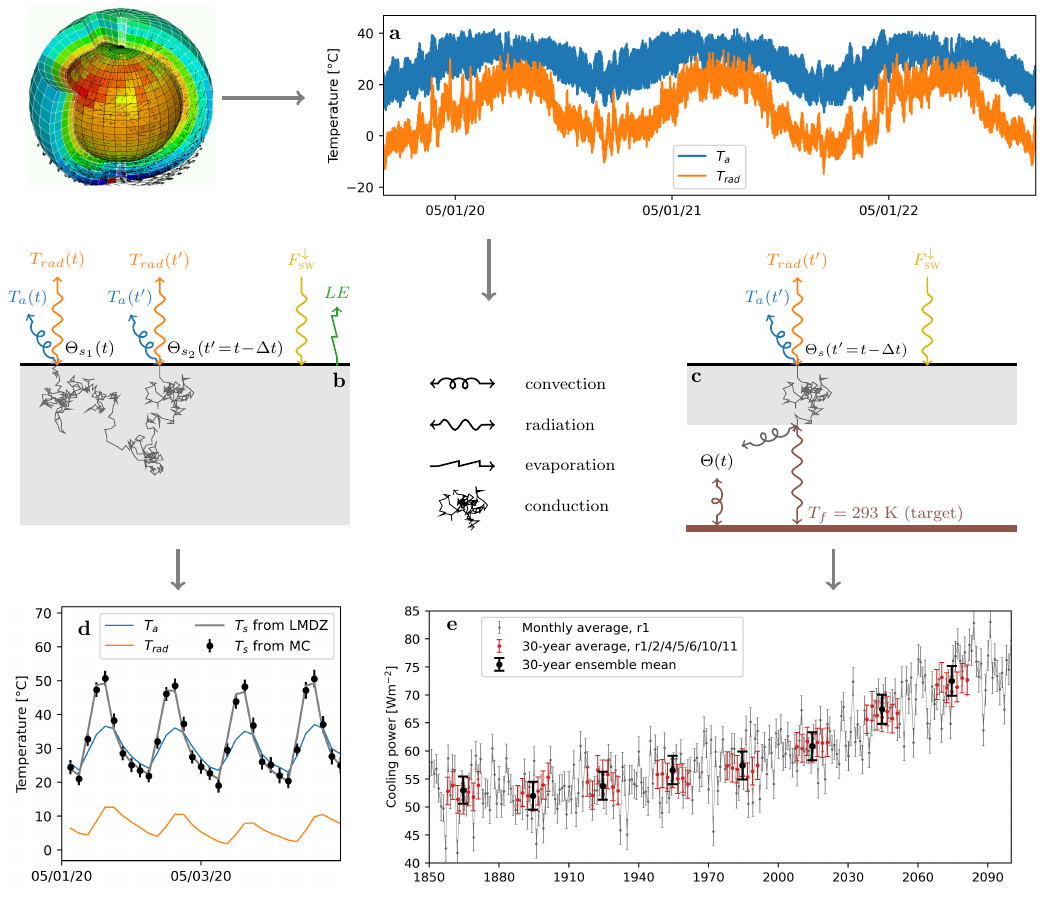}
  \caption{Time-varying meteorological conditions are used as inputs and
  parameters in path-integral heat transfer models. \sfiglab{a}, the air
  temperature at 2 meters above the surface ($\tatm$) and the atmospheric
  brightness temperature ($\trad$) issued from a climate change simulation
  performed with the IPSL-CM6A-LR global model \cite{Bouc:20}, available at a
  3~hour frequency over 250 years. The variables retrieved from the climate archive
  are: $\tatm$; the downwelling longwave ($\flw$, used to compute $\trad$) and
  swortwave ($\fsw$) radiative fluxes at the surface; the sensible ($\sens$)
  and latent ($\flat$) turbulent heat fluxes; and the surface temperature
  $\tsrf$. $\sens$ and $\tsrf$ are used to compute a convective exchange
  coefficient $h=H/(\tsrf-\tatm)$. $\flat$ and $\fsw$ are imposed fluxes. The
  data correspond to a gridpoint in Sahel. \sfiglab{b-c}, Random path
  representation of the heat transfer models used to estimate: \sfiglab{d}, the
  surface temperature of a homogeneous soil of thermal inertia
  1500~J~m$^{-2}$~s$^{-1/2}$~K, and \sfiglab{e}, the air-conditioning power to
  maintain a simplified room's floor at 293~K. \sfiglab{d}, instantaneous
  temperatures every 3~h during four days: Monte Carlo estimates of $\tsrf$
  (black dots) and $\tsrf$, $\tatm$ and $\trad$ from the climate archive (gray,
  blue and orange lines). \sfiglab{e}, May averages of air-conditioning power
  from 1850 to 2100: every year (gray dots); averaged over 30 years (red dots),
  each red dot corresponds to a different member of an ensemble simulation
  \cite{Bonn:21Nat}; averaged over 30 years and over the ensemble members
  (black dots). Dots and error bars in \sfiglab{d} and \sfiglab{e} correspond
  to Monte Carlo estimates based on 30k paths and their associated 99.7\%
  confidence interval.
  }
  \label{fig:climat}
\end{figure*}

To inform adaptation strategies aiming at minimizing urban heat stress and
energy consumption, the performances of new construction materials, building
and city designs need to be assessed in realistic environmental conditions
representative of future climates, at various spatio-temporal scales, with
unconstrained amounts of geometric details. The properties of Monte Carlo probe
computations open an avenue for such computations by offering the possibility
to sample weather conditions in addition to the other dimensions of the problem
\cite{farges_life-time_2015}. \fig{climat} illustrates in a simple geometry how
the approach allows to integrate the large temporal scale factors from the
meteorological-process scale to the climate scale in a relevant multi-physics
calculation. Outputs of an ensemble of global climate simulations of 250 years
each are used, available at a frequency of 3 hours \cite{Bonn:21Nat}.

In the first example, the estimated quantity is the temperature at the surface
of an homogeneous soil at a given time (\sfig{climat}{b,d}). Paths start at the
soil's surface and travel through the system and backward in time. Each time
the path encounters the surface, solar and evaporative heat fluxes are added to
the Monte Carlo weight and the flux continuity equation gives the probability
that the path goes into the atmosphere by convection or infrared thermal
radiation (the path ends with outcome $\tatm$ or $\trad$ respectively) or
penetrates the ground by conduction.

Transient conduction is simulated using diffusive random walks. At each step,
the duration associated with the step length is sampled from an exponential law
parameterized by the material inertia, as per the Green first-passage times
distribution function. The walk goes on until the initial condition (at year
1850 here) or the surface is reached. Encounters with the surface will
therefore happen at different times depending on the random duration of the
walk. It is only at these times that the meteorological data need to be
accessed. The time dimension is thus sampled based on the physical properties
of the system: the longest paths will go back farther in time if the thermal
inertia of the ground is larger. Moreover, the influence of the meteorology
onto the surface temperature is sampled according to the detailed
meteorological processes. For instance, the probability for a path to end up in
the air by convection is generally smaller during nightime than during daytime
due to a smaller value of the convective exchange coefficient.

The second example (\sfig{climat}{c,e}) is the computation of the
air-conditioning power needed to maintain a room's floor temperature to a
setpoint of 20°C. In the upper part of the geometry (atmosphere + roof) the
model is the same as in the previous example except that latent heat fluxes are
neglected. In the lower part, the temperature at the bottom of the slab roof
(i.e., the ceiling) is coupled to that of the (perfectly-mixed) room's air by
convection and to that of the floor by radiation. The air-conditioning power is
calculated as the net heat flux between the floor and the system.

Using double randomization, a single Monte Carlo simulation is used to estimate
the power not at a particular time but on average over a given period. This is
achieved by sampling a different starting time at the beginning of each path.
Because this additional sampling does not increase the samples variance, the
same number of paths and hence the same computing time is needed to reach a 1\%
accuracy as for a single-time estimate. {Additionally} sampling the members
of an ensemble of climate simulations gives an estimate of the power averaged
over the ensemble of simulated meteorological stories, again at the same cost
as for a single-member single-time computation.

This very preliminary computation obviously suffers from several limitations.
The building geometry is oversimplified compared to \fig{benjamin}, as is the
treatment of atmospheric radiation compared to \fig{nuages}. Firstly, the codes
that have been used to produce these images (stardis and htrdr) still need to
be coupled together and interfaced with the climate data. Important work
remains to produce geometric data at the required format in a way that is
flexible enough to allow simple user modifications. Secondly, the temperature
of the near-surface air (used to compute turbulent or convective fluxes) and
the downward thermal radiative fluxes are unaffected by the temperature of the
building which prevents the representation of the UHI effect. Thirdly, in
contrast to Urban Canopy Models or Obstacle Resolving Models that are often
based on the resolution of fluid dynamics but struggle to integrate a full
description of the thermal transfers in the buildings, the Monte Carlo methods
easily solve the physical and geometrical complexity of the thermal and
radiative transfer in the buildings but struggle to solve the atmospheric flow.

A research program is currently funded by the French National Agency for
Research to overcome these limitations. The perturbation of the air
temperature above the buildings will indirectly be taken into account by
pursuing the paths in the atmosphere through turbulent, convective updrafts or
advective motions. The path will end with the air temperature from the model
only once the path is outside the city's footprint, thereby representing the
urban heat island. On the other hand, the Monte Carlo path-tracing algorithm
that solves the thermal exchanges in the city will be coupled to an Obstacle
Resolving Model of the flow in the city. For this coupling, the temperature has
to be estimated at all the buildings interface with the atmosphere hence
alternatives to the classical probe computation are studied to accelerate the
computation such as the Symbolic Monte Carlo methods
\cite{galtier_symbolic_2017}. Only relatively short simulations will be
produced this way, to serve as a reference to benchmark faster models; this
strategy was already proven successful for the development of cloud
parameterizations, using explicit LES as a reference \cite{Couv:21}.

Note that for adaptation issues, such computations should rely on global
coupled climate models, the only models able to simulate the thousands of years
of meteorological evolution that are required to achieve equilibrium and
simulate climate change. The representation of convective and cloud processes
in these models has strongly improved in the last decades thanks to recent
advances in parameterizations and model tuning \cite{Hour:21}, although much work
remains to reduce the uncertainties associated with the representation of these
processes. Because of persisting biases in global models and of their rather
coarse resolution, statistical or dynamical downscaling, for instance using
regional climate models, might prove necessary to better account for local
constraints or detailed processes such as topography or the radiative effect of
geometrically complex clouds. This opens exciting questions that are yet to be
investigated.

A recurrent challenge of climate services is that climate records and
simulations represent huge amounts of data from which information relevant to
the user's needs has to be extracted, preferably in a comprehensible, flexible
way so that end users can be in full responsibility of their work. Here the
idea is to run probe computations taking the full history of the atmospheric
column as an input, provided by global or regional climate simulations. No
pre-treatment of the data is needed: through path integrals, the physics define
a relevant way to aggregate the climate data. Flexibility is ensured by the
fact that each quantity to be estimated will be associated with its own
path-integral, its own tailored ``data mining'' procedure. This implies that
the data output from climate simulations must be made entirely available, which
is in line with the open-science (open-source and open-data) philosophy. It can
of course raise practical issues, slowing the computations down if data access
is not managed efficiently. Packaging software in containers to be run on
servers where the data are stored, or downloading a single column of the full
history of climate simulations beforehand as was done for the simple
illustration of \fig{climat}, might be adequate solutions.

An important aspect of our proposition is the empowerment of users. As for 3D
animation, this framework is intended to set free the users who could add any
details to the building designs and materials without fearing consequences on
the numerical cost of their choices, thanks to the independence between the
algorithms and the data description. Moreover, we believe that the paths convey
meaningful images that have the potential to enlighten scientists and
non-scientists with intuitive understanding of the physical processes at stake
and their interactions. The theoretical framework and associated numerical
tools provide more than numbers; they provide insights into the questions at
the heart of the energy transition.

\section*{Authors contribution} {All authors have contributed to the
formulation and structuration of the ideas exposed in the article, as well as
to internal review of the origin manuscript. In addition, N.V. and F.H.  led
the manuscript writing;  N.V., F.H., L.d'A., V.E., V.F. and B.P.  contributed
to the results and figures; V.M. and R.S.  contributed to writing the
introduction and the content of the box; R.F. and S.B. contributed to the
bibliography.}

\section*{Data availibility statement}
The models description, source codes and data that were used to produce the
results discussed in this publication are available at
\url{https://www.lmd.jussieu.fr/~nvillefranque/pages/teapot_city.html} and
DOI:10.14768/c55c1f3d-5223-48ef-8713-593414adfbdd. {Most of this work is based
on open-source software developed, maintained and distributed under free
license by Méso-Star
\url{https://www.meso-star.com/projects/misc/about-en.html}.} All other data
needed to evaluate the conclusions in the paper are present in the paper.

\section*{Competing Interests}
The authors declare that they have no competing interests.

\section*{Acknowledgment}

{Beyond the authors of the present article, the advances on Monte Carlo
methods and the results reported here owe much to the EDStar research
consortium. We thank Igor Roffiac for inspiration, and Sébastien Hourdin whose
work inspired the simple example of the 2D-building.} We thank Fleur Couvreux
and Jean-Yves Grandpeix for their important role as alpha-readers and their
useful remarks on the original version of this work. This work received
financial support by the French National Agency for Research (ANR project MC2,
ANR-21-CE46-0013 and ANR project MCG-RAD ANR-18-CE46-0012) and by the French
Agency for Ecological Transition (ADEME project MODRADURB-1917C001). The
simulations run to produce the pictures of the animation movie ``A teapot in
the city under cumulus clouds'' were granted access to the HPC resources of
IDRIS under the allocation gencmip6 attributed by GENCI (Grand Equipment
National de Calcul Intensif). {For the availability of climate simulations
output, we benefited from the ESPRI computing and data center
(https://mesocentre.ipsl.fr), which is supported by CNRS, Sorbonne Université,
Ecole Polytechnique, and CNES. The simulations themseleves were performed by
the IPSL-Climate Modeling Center.}

\end{document}